\documentclass[aps,pre,twocolumn,superscriptaddress,amsmath,amssymb,showpacs,showkeys]{revtex4}
\usepackage{graphicx}

\newcommand{\beq}{\begin{equation}}
\newcommand{\eeq}{\end{equation}}

\newcommand{\caL}{{\mathcal L}}

\newcommand{\opunit}{\text{1}\kern-0.22em\text{l}}

\newcommand{\ie}{i.e.\;}

\usepackage{natbib}
\begin{document}
\bibliographystyle{plainnat}

\title{Amplification of compressional MHD waves in systems with forced entropy oscillations}

\author{Bidzina M. Shergelashvili}
\email{Bidzina.Shergelashvili@fys.kuleuven.be}%
\affiliation{Instituut voor Theoretische Fysica,
K.U.Leuven,
Celestijnenlaan 200 D, B-3001, Leuven}%
\affiliation{Centre for Plasma Astrophysics, K.U.Leuven,
Celestijnenlaan 200 B,
B-3001, Leuven}%
\affiliation{Georgian National Astrophysical Observatory, 2a,
Kazbegi Ave., 0160 Tbilisi, Georgia}%
\author{Christian Maes}
\affiliation{Instituut voor Theoretische Fysica, K.U.Leuven,
Celestijnenlaan 200 D, B-3001, Leuven}%
\author{Stefaan Poedts}
\email{Stefaan.Poedts@wis.kuleuven.be} \affiliation{Centre for
Plasma Astrophysics, K.U.Leuven,
Celestijnenlaan 200 B, B-3001, Leuven}%
\author{Teimuraz V. Zaqarashvili}
\email{temury@genao.org}%
\affiliation{Georgian National Astrophysical Observatory, 2a,
Kazbegi Ave., 0160 Tbilisi, Georgia}%
\date{\today}

\begin{abstract}
The propagation of compressional MHD waves is studied for an
externally driven system. It is assumed that the combined action
of the external sources and sinks of the entropy results in the
harmonic oscillation of the entropy (and temperature) in the
system. It is found that with the appropriate resonant conditions
fast and slow waves get amplified due to the phenomenon of
parametric resonance. Besides, it is shown that the considered
waves are mutually coupled as a consequence of the nonequilibrium
state of the background medium. The coupling is strongest when the
plasma  $\beta \approx 1$. The proposed formalism is sufficiently
general and can be applied for many dynamical systems, both under
terrestrial and astrophysical conditions.

\end{abstract}


\maketitle


\section{Introduction\label{intro}}

Entropy plays a central role in the thermodynamic properties of
fluids and plasmas that consist of an enormous number of
particles. In general terms and as a function of energy, volume,
etc., the entropy is a measure of disorder as it counts the
multitude of microscopic realizations for some given manifest
condition. That disorder is realized in the chaotic motions of the
system entities and in the dissipative character of transport
processes. Under normal circumstances, when the system is left to
itself, it evolves to the state of maximum entropy which is
compatible with the imposed conditions. That relaxation to a
so-called equilibrium is one realization of the direction of time
and it could be applied to any closed system or to the universe as
a whole for that matter.

Open systems, however, can be found and for a long time in a state
of low entropy. In that case, these systems reach a steady state
regime in which they are constantly driven out of equilibrium.  An
important topic for investigation is the study of various
dissipative effects and their relation with fluctuation and
transport phenomena.  It is of great interest to understand if and
how the low entropy is realized in some type of macroscopic
pattern or structure, i.e.\ a collective and ordered motion which
stands in great contrast with the chaotic motion at a microscopic
level. Obviously, macroscopic order is also possible at
equilibrium, e.g.\ at low temperature or at high pressure, but in
the present paper we aim to describe a nonequilibrium effect in
which a nontrivial macroscopic effect is {\it generated} in time.
More specificly, our study involves the amplification of wave
motions in a driven system.

In order to organize the chaotic motions of the particles into
collective motions exhibiting long range or long term non-trivial
correlations, the effort of some external driver is required. This
external source supports the system by delivering work or
exchanging heat. Secondly, the system must react to the driver in
such a way that the interaction between the different structures
is sufficiently rich to select a nontrivial mode of operation. The
specifics needed of that nonlinear dynamics remain more vague as
today there is no unifying understanding of morphogenesis.  In the
present paper, the mechanism at work is that of parametric
resonance, as we will explain below.

We consider  compressional magnetohydrodynamic (MHD) waves
(structures) propagating in a magnetized plasma (the system
consisting of chaotically moving ions and electrons), which is
driven by external heat reservoir(s). Mechanisms of excitation,
conversion, amplification etc.\ of different wave modes are of
special importance for many terrestrial and astrophysical systems.
The waves represent effective transmitters of the energy and of
the information present in the background system. There are many
studies concerning the wave propagation in driven systems. One of
the many examples is shear flow. In a shear flow, an inhomogeneous
velocity profile is formed as a consequence of the action of the
external force. The flow is steady and the system is in a
stationary nonequilibrium state.  In particular, the work done by
the external force does not increase the system's energy. On the
contrary, the delivered energy is dissipated and, more
interestingly, different wave modes exchange energy with each
other and are coupled by the flow , see for instance
\cite{chartsi96,gogo2004} and references therein, and
\cite{bufar1992,tre2005,tre1993} for more applications in the
context of the nonmodal and pseudospectral analysis of such
systems; the role of shear flows in the formation of the observed
$p$-mode power spectra is discussed in \citep{shergpm05} and its
role in the heating of the magnetically governed solar corona is
found in \cite{shergself06}.

One of the essential mechanical properties of fluids and plasmas
is the possibility of parametric resonance, which might occur in a
system sustaining different kind of periodic in time (or space)
processes. Recently, the ability of different types of MHD waves
to be involved in so-called swing (or parametric) interactions,
has been studied e.g. in \cite{zaq2002b,zaq2006a}, including
spatially inhomogeneous backgrounds \cite{shergsw05}. However, in
all these cases the problem was concerned with non-driven systems,
with waves representing the perturbations above some equilibrium
configurations of the plasma. When an external periodic force on
the compressional fast and slow magnetosonic waves was considered
\cite{zaqolb2002}, nevertheless the external force applied to the
system was treated as adiabatic, \ie, it did not make any
contribution into the variation of entropy (or temperature).
Therefore, even with this action the characteristic sound speed
remained constant and the parametric action of this force on the
waves had been achieved by the periodic oscillation of the modal
wavelength only.

In the present paper we want to understand and to clarify a
similar process when the external driver is a cause of entropy (or
temperature) variation. Some efforts have already been taken to
extend the model in this direction before. For instance, studies
have been conducted on the parametric amplification of acoustic
waves \cite{berktay1965,shapiro1975,zarembo1975} also including
systems that are exposed to periodic variations of external
electric (piezoelectric medium) \cite{mansfeld2003} or magnetic
(magnetostrictive solids) \cite{voinovich2004} fields. These
variations indeed give birth to oscillations of the phase speeds
of the waves leading to the parametric amplification of them under
certain resonant conditions. In spite of these investigations, a
more general theory of parametric amplification of fast and slow
magneto-sonic waves in systems with periodic entropy production
has not been developed. We address this issue in a systematic
manner to develop a unified formalism for the qualitative and
quantitative treatment of the problem.

The paper is organized in the following way: in the next section
\ref{bmodel} we develop a basic MHD model and specify the
properties of the background nonequilibrium state; Further, in
section \ref{fswaves} we study the propagation of fast and slow
magnetosonic waves in the system (also we give a discussions about
the hydrodynamic limit); And in section \ref{concl} we give some
conclusive remarks.

\section{Basic model\label{bmodel}}

The aim of the present paper is to study properties of the
compressional fast and slow magneto-sonic waves in a magnetized
single component plasma, which is driven by one or more external
reservoirs that bring it out of  thermodynamic equilibrium. We can
write the full set of MHD equations to describe the dynamics of
such plasmas in the following way:
\beq \label{basica}%
\frac{D\rho }{Dt}+\rho (\mathbf{\nabla}\mathbf{\cdot
}\mathbf{V})=0,%
\eeq%
\beq \label{basicb}%
\rho \frac{D\mathbf{V}}{Dt}=-\mathbf{\nabla}\left[ p+\frac{B^{2}}{8\pi }\right] +\frac{%
\left( \mathbf{B}\cdot \mathbf{\nabla}\right) \mathbf{B}}{4\pi } +\nu \Delta \mathbf{V},%
\eeq%
\beq\label{basicc}%
\frac{D\mathbf{B}}{Dt}=\left( \mathbf{B}\cdot \mathbf{\nabla}\right) \mathbf{V}-\mathbf{B}\left( \mathbf{%
\nabla}\cdot \mathbf{V}\right)+\eta \Delta \mathbf{B},%
\eeq%
\beq\label{basicd}%
 \frac{Dp}{Dt}-\frac{\gamma p}{\rho }\frac{D\rho
}{Dt}=(\gamma
-1)\caL,%
\eeq%
where all the quantities have their usual meaning and where
$D/Dt=\partial /\partial t+\left( \mathbf{V}\cdot
\mathbf{\nabla}\right)$ denotes the convective derivative. The
symbols $\nu$ and $\eta$ denote the coefficients of viscosity and
magnetic diffusion, respectively. However, we write these
dissipative terms in the equations just for the sake of
generality; as will be shown below, the effects of viscosity and
resistivity can be omitted in this consideration. Such a more
general set-up is necessary to indicate that, in general, the
mentioned effects can also give rise to a contribution into the
source term, $\caL$ (i.e., in the production of entropy), standing
on the right-hand side (RHS) of Eq.~(\ref{basicd}).%

\subsection{The background state}

Before we turn to the study of the wave properties in the
considered nonequilibrium system, it is convenient to specify the
background state. In fact, this is desirable because usually, in
the literature, the wave processes are considered to be adiabatic,
i.e.\ the action of non-adiabatic processes, leading to the
generation and/or transport of the entropy within the system and
its surroundings, is usually omitted. This approach implies a
definite separation between two characteristic timescales, viz.\
(i)~that of the non-adiabatic variation of the entropy and
(ii)~the timescale of collective (macroscopic, mechanical)
phenomena, for instance the wave and the oscillatory motions. From
the structure of the basic set of equations
(\ref{basica})-(\ref{basicd}), it is evident that we consider also
non-adiabatic processes, which can bring the system away from the
thermodynamical equilibrium. In general, Eq.~(\ref{basicd})
represents an entropy balance equation
of the form:%
\beq\label{entrop}%
\rho T \frac{DS}{Dt}=\caL,
\eeq%
where $T$ and $S$ denote the temperature and entropy per unit
volume, respectively. The source term in this equation represents
the sum of all the sources and sinks of the local entropy. As is
well known from the characteristic analysis, when one considers only
the adiabatic motions of a plasma or a fluid ($\caL\approx 0$, or in
other words, when the background state is in exact equilibrium) then
any small perturbations of the entropy are aperiodic (often referred
to as entropy waves/modes) and they are decoupled from the other
fundamental eigenmodes of the system (which are of oscillatory
nature). Hence, in that case, the local entropy remains unchanged as
these other eigenmodes propagate.

When the system is open, however, the local processes can be
considered as part of a larger system which itself is closed but
still enables local variations (even decreasing) of the entropy.
When the characteristic timescales of the considered processes are
comparable with the timescale of the entropy variation, the source
term in Eq.~(\ref{entrop}) can be significant and, thus, should
not be neglected in that case. In this situation, the entropy
variations can be coupled with the other modes of collective
motion which are sustained by the system. If the temporal
variation of the entropy source is harmonic, then one can refer to
the entropy deviations from its equilibrium value as `forced
entropy modes' (or oscillations).

In general, one can consider a static equilibrium characterized by
the entropy $S_{00}$, with an homogeneous pressure
$p_{00}=\text{const}$ and density $\rho_{00}=\text{const}$. In
this equilibrium, the source term vanishes, i.e.\ $\caL _{00}=0$.
Further, we introduce a small time-dependent deviation from the
equilibrium, so that $S_0=S_{00}+S_{01}(t)$. This deviation is due
to the presence of external reservoirs which exchange entropy with
the system leading to a finite time-dependent perturbation
modelled by the term $\caL_{01}(t)$. The physical quantities
$p_0=p_{00}+p_{01}(t)$ and $\rho_0=\rho_{00}+\rho_{01}(t)$ then
satisfy the following equation:
\beq\label{entpert}%
 \frac{Dp_0}{Dt}-\frac{\gamma p_0}{\rho_0}\frac{D\rho _0
}{Dt}=(\gamma
-1)\caL _{01}(t).%
\eeq%
This equation, in combination with the appropriate equations from
the basic set (\ref{basica})-(\ref{basicc}), implies that the
entropy variations in general are not decoupled from the motions
related to the fundamental modes in the system. Moreover, under
certain conditions (which we will specify below) these equations can
describe forced oscillations.\footnote{Here, it should be mentioned
that the forcing of the oscillations of the system can be achieved
not only by the generation and the absorption of heat, but also
through the application of periodic external forces, which would
work on the considered system. This remark might be of importance
also for experimental purposes, in case one would want to realize
similar processes under terrestrial, laboratory conditions.}

Now let us consider the source term $\caL _{01}(t)$ in more detail.
There are different (transport) processes which may contribute to
this term, which formally could be attributed to two classes, viz.\
(i)~the processes of the entropy production (such as viscous and
resistive dissipation, thermonuclear reactions, chemicals reactions
etc.), and (ii)~the transport of entropy (via thermal conduction,
radiation, diffusion, etc.). Following the representation given by
\citet{grootmaz}, this term can be represented in its most general
form as:
\beq\label{source}%
\caL _{0} (t) =T_0 \left (-\left( \mathbf{\nabla}\cdot
\mathbf{J}_{s0} \right) + \sigma_{0}\right ).
\eeq%
Here, $\mathbf{J}_{s0}$ denotes the entropy current, and
$\sigma_0$ denotes the entropy production. We should note that
$\mathbf{J}_{s}=\mathbf{J}_{s,\text{tot}}-\rho S \mathbf{V}$,
i.e.\ $\mathbf{J}_{s}$ represents only the microscopic current of
the entropy, while the macroscopic transport $\rho S \mathbf{V}$
is excluded \cite{grootmaz}.

Further, we represent both these quantities as the sum of a
constant and a variable part, i.e.\
$\sigma_0=\sigma_{00}+\sigma_{01}(t)$, and
$\mathbf{J}_{s0}=\mathbf{J}_{s00}+\mathbf{J}_{s01}(1)$. These are
the general representations of the entropy production and the
entropy current. In this paper, we are interested to study systems
in which the constant parts of these two effects compensate each
other, supporting the development of thermodynamic equilibrium
when entropy variations in time are absent:
\beq\label{constent}%
\caL _{00}=T_0 \left (-\left( \mathbf{\nabla}\cdot
\mathbf{J}_{s00} \right) + \sigma_{00}\right )=0,
\eeq%
and the remaining part of the source term, $\caL _{01}$, depends on
time in a harmonic way:
\beq\label{harmonent}%
\caL _{01}=T_0 \left (-\left( \mathbf{\nabla}\cdot
\mathbf{J}_{s01} \right) + \sigma_{01}\right )=-\alpha \sin{\Omega
t},
\eeq%
where $\Omega$ is the frequency of the entropy variation in time
and $\alpha$ denotes a constant amplitude. This amplitude is
rather small as we consider only small oscillations around
thermodynamic equilibrium. Naturally, $\alpha$ might depend also
on the spatial coordinates leading to a spatial variation of the
thermodynamic quantities as well. However, we here assume that
this variation is rather weak. In other words, the characteristic
length of this variation is assumed to be larger than the system
size and the action of the external reservoirs thus causes an
oscillation of the entropy everywhere within the system at once.
Consequently, in the equations governing the dynamics of the
background state, we replace the convective derivatives by partial
derivatives.

Now, Equation~(\ref{entpert}) reveals the fact that the considered
variation of the entropy can result in a variation of the other
physical quantities, for instance the plasma pressure, density,
temperature, velocity, etc. The purpose of the present work is to
discover the basic nature of the resonant processes which may take
place in the described dynamical system. And, therefore, and for
the sake of further simplification, we assume that the entropy
oscillation leads only to an oscillation of the pressure and the
temperature, while the plasma density remains constant (i.e.\
$\rho_{01}(t) \equiv 0$). This is possible in any configuration,
which admits only a very limited ability for the system volume to
vary significantly or when the system is not able to alter its
volume at all globally in the background (isochoric process),
while small fluctuations of the density are still possible. With
this assumption, Eq.~(\ref{entpert}) takes the form:
\begin{equation}\label{entpert2}
\frac{\partial p_{0}}{\partial t}=-\alpha _1 \sin (\Omega t),
\end{equation}
where, $\alpha _1=(\gamma -1)\alpha$. This equation can be
integrated to obtain $p_0=p_{00}+p_{*}\cos(\Omega t)$ (here,
$p_{*}=\alpha_1/\Omega$), which immediately implies that the
characteristic sound speed of the system is a harmonic function of
time as well
\beq\label{sound}%
C_s ^2=\frac{\gamma p_0}{\rho _0}=C_{s0} ^2(1+\delta \cos(\Omega
t)),
\eeq%
where $C_{s0}$ and $\delta$ are constant parameters.

One might be interested in an example of such a configuration where
this kind of situation can be realized. Consider, for instance, a
static ($\mathbf{U}_0=0$) system with thermal conduction driven by
the heat flux through the boundaries. For such a system one can then
write down the following Fourier's equation:%
\beq\label{fouriertemp}%
\rho C_v \frac{\partial T}{\partial t} = \lambda \Delta T,
\eeq%
where $C_v$ denotes the specific heat at constant volume
(density), $\lambda$ is the coefficient of the thermal conduction,
and $\Delta$ denotes the Laplacian \cite{grootmaz}. This equation
is one realization of Eq.~(\ref{basicd}). In the appendix, we also
show the relation between the coefficient of thermal conduction
and the local production and transport of entropy. So, if the
boundary condition is a harmonic function of time, then the
temperature profile within the system would be also oscillating
representing an example of the realization of a set-up given by
Eq.~(\ref{entpert2}).

Consider the case when the temperature at the boundary $x=0$ is a
periodic function of time:
\beq%
 T_0=T_{00}+T_{01}e^{-i\omega t},\,\, x=0.%
\eeq%
The distribution of the temperature within the system then has the
same time dependence, i.e.\ $\exp{-i\omega t}$. The real
temperature distribution will then be \citep[cf.][]{lanlifsh}
\beq\label{tempmode}%
T_0=T_{00}+T_{01}\exp \left (-x\sqrt{{\omega \rho C_v}\over
{2\lambda}} \right )\exp \left (ix\sqrt{{\omega \rho C_v}\over
{2\lambda}} -i\omega t \right ).
\eeq%
Therefore, it is seen that the temperature oscillation propagates
in the form of damped thermal waves. Here, one can avoid the
spatial dependence in a similar way as it was done by Lorenz to
study the properties of the Rayleigh-B\'enard convection cells
\cite{lorenz1963}. However, this can be done only when the
wavelength of the fundamental harmonic of the temperature waves
given by Eq.~(\ref{tempmode}) is larger than the system size (\ie
when either the coefficient $\lambda$ is large or the density (or
oscillation frequency) is small enough).

There is also another way to achieve the variability of the
background temperature: if we have an external source of heat $Q$
(for example, in the laboratory situation, this source can be the
heating by electric currents or lasers), then
Eq.~(\ref{fouriertemp}) can be rewritten as
\beq%
\rho C_v \frac{\partial T}{\partial t} = \lambda \Delta T + Q.
\eeq%
Now, if we take $Q$ as periodic function of time and homogeneous in
space, then the temperature (pressure) may have a periodic time
dependence as well, thus leading to Eqs.~(\ref{entpert2}) and
(\ref{sound}).

\section{Fast and slow magneto-sonic waves}\label{fswaves}

In this section, we derive the dynamical equations governing the
propagation of the linear eigenwave modes of the system. For this
purpose we linearize the full set of MHD equations over the
background nonequilibrium state outlined in the previous section.
In principle, the process we are addressing here represents a
weakly non-linear action of the entropy oscillations, forced by
the combined effect of all sources and sinks of entropy upon the
eigenmodes of the system. At this stage, we treat the problem in
two spatial dimensions only to avoid unnecessary mathematical
complications. The plasma is embedded into a uniform magnetic
field directed in the $z$-direction. The $x$-axis is assumed to
cover the direction across the magnetic field. We know that, in
this particular case, the system sustains only two kinds of
fundamental modes, viz.\ fast and slow magneto-sonic waves. The
linearized set of equations then reads:
\beq\label{lineara}%
\frac{\partial \rho ^{\prime }}{\partial t}+\rho _{0}\left(
\frac{\partial u_{x}}{\partial x}+\frac{\partial u_{z}}{\partial
z}\right) =0,%
\eeq%
\beq\label{linearb}%
\frac{\partial u_{x}}{\partial t}=-\frac{1}{\rho _{0}}\frac{\partial }{%
\partial x}\left[ p^{\prime }+\frac{B_{0}b_{z}}{4\pi }\right] +\frac{B_{0}}{%
4\pi \rho _{0}}\frac{\partial b_{x}}{\partial z}+\nu\Delta u_{x},%
\eeq%
\beq\label{linearc}%
\frac{\partial u_{z}}{\partial t}=-\frac{1}{\rho
_{0}}\frac{\partial p^{\prime }}{\partial z}+\nu\Delta u_{z},%
\eeq%
\beq\label{lineard}%
\frac{\partial b_{x}}{\partial t}=B_{0}\frac{\partial
u_{x}}{\partial z}+\eta\Delta b_{x},%
\eeq%
\beq\label{lineare}%
\frac{\partial b_{z}}{\partial t}=-B_{0}\frac{\partial
u_{x}}{\partial x}+\eta\Delta b_{z},%
\eeq%
\beq\label{linearf}%
\frac{\partial p^{\prime }}{\partial t}-\frac{\gamma p_{0}}{\rho _{0}}\frac{%
\partial \rho ^{\prime }}{\partial t}=\caL ^{\prime},
\eeq%
where the perturbed vector fields of velocity and magnetic field are
written by lowercase symbols, while the scalar fields (pressure and
density) are denoted by symbols with primes. Further, we assume that
the wavelength of the waves that are considered here are larger than
the characteristic dissipation scales in the system. Therefore, we
can neglect the terms corresponding to the viscosity $\nu\Delta
u_{x},\nu\Delta u_{z}\approx 0$ and the magnetic diffusion
$\eta\Delta b_{x},\eta\Delta b_{z} \approx 0$. Consequently, we
assume that the perturbation of the source term in
Eq.~(\ref{linearf}) vanishes, i.e.\ $\caL ^{\prime}\approx 0$, as it
represents a contribution in the entropy variation from the
dissipation of the perturbations. In other words, we consider the
situation when the energy of the external reservoirs driving the
system is much larger than the energy of the waves and a back
reaction of the latter on the system is negligibly small. Yet, the
effect of the variable background is represented by the variable
pressure or sound speed.

The governing set of equations (\ref{lineara})-(\ref{linearf}) is
homogeneous w.~r.~t.\ the spatial coordinates. Hence, we can apply
a Fourier analysis by representing all physical quantities as:
\beq\label{fourier}%
f=\int \int \hat{f}(k_x,k_z,t)\exp i(k_xx+k_zz) dk_xdk_z,
\eeq%
and with the above mentioned assumptions we obtain a set of two
second order ODEs:
\beq\label{wavebasux}%
\frac{d^{2}\hat{u}_{x}}{dt^{2}}+c_{11}\hat{u}_{x}+c_{12}\hat{u}_{z}=0,%
\eeq%
\beq\label{wavebasuz}%
\frac{d^{2}\hat{u}_{z}}{dt^{2}}+c_{12}\hat{u}_{x}+c_{22}\hat{u}_{z}=0,%
\eeq%
where $c_{11}=C_{s}^{2}(t)k_x^{2}+V_{A}^{2} k^{2}$,
$c_{12}=C_{s}^{2}(t)k_x k_z$, $c_{22}=C_{s}^{2}(t)k_z^{2}$ (note
that here, $k^2=k_x ^2+k_z ^2$ and $V_A=B_0/\sqrt{4\pi \rho _0}$
corresponds to the Alfv\'en speed). These equations govern the
evolution of the oscillatory motions.  Before we make a rigorous
analysis of this system we consider the hydrodynamic case, i.e.\
when $B_0=0$.

\subsection{The hydrodynamic case}\label{hydrolim}

\begin{figure*}
  \includegraphics[scale=0.9]{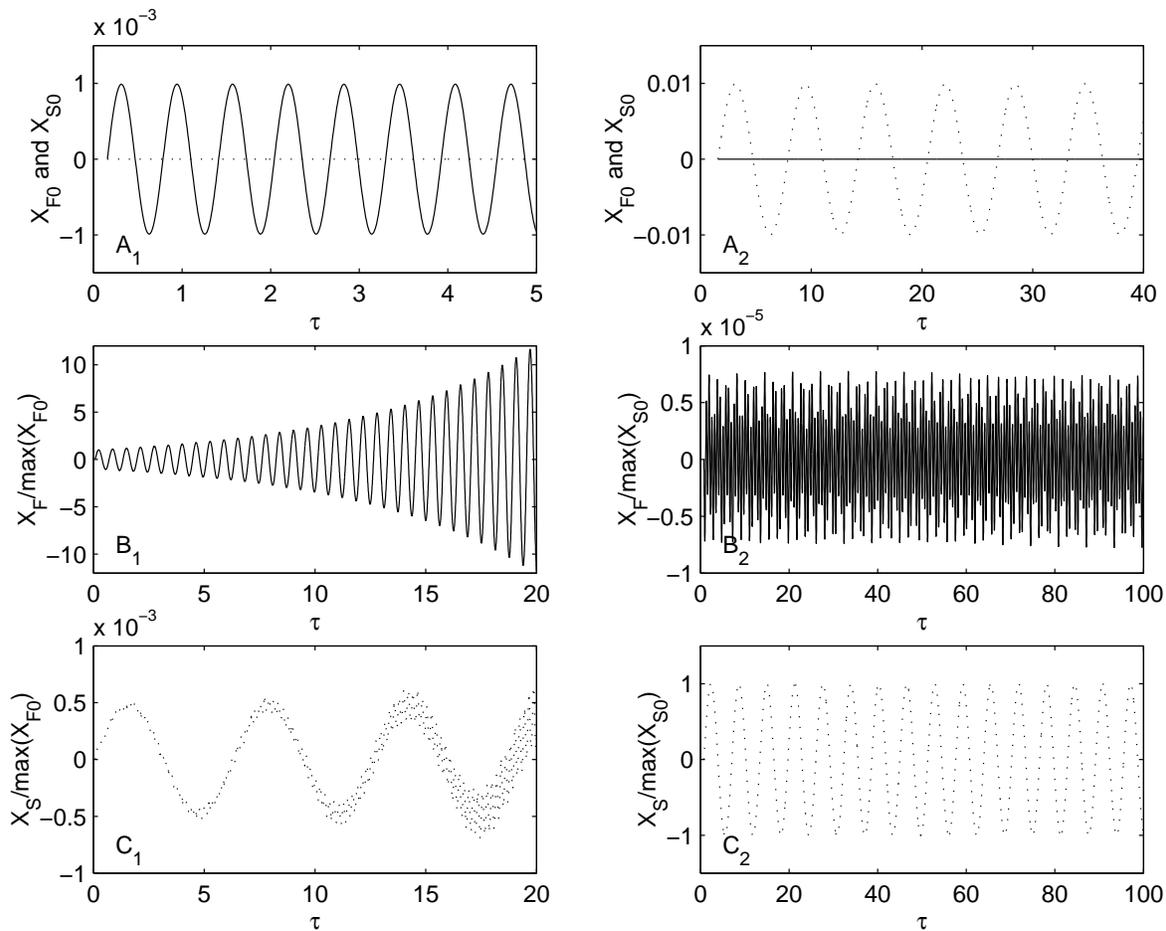}\\
  \caption{\label{figxlarge} $X_{\text{F}}$ (solid lines) and $X_{\text{F}}$ (dotted lines)
  plotted against time $\tau$.}
\end{figure*}

When the magnetic field is absent, the system
(\ref{wavebasux})-(\ref{wavebasuz}) can be rewritten in the form
of a single wave equation. Introducing the new variable:
\beq\label{xa}%
X_{\text{ac}}=k_x\hat{u}_{x}+k_z\hat{u}_{x},
\eeq%
we can write:
\beq\label{waveac}%
\ddot{X}_{\text{ac}}+C_{s0} ^2k^2(1+\delta \cos(\Omega
t))X_{\text{ac}}=0,
\eeq%
where the number of dots above the variable indicates the order of
time-derivative. This reduction to a single wave equation
represents the fact that there are only compressional acoustic
waves in the hydrodynamic limit. It is evident that
Eq.~(\ref{waveac}) is a Mathieu-type equation and, thus, an exact
analytical solution is available. The solution has a resonant
nature when:
\beq\label{resac}%
\Omega \approx 2C_{s0} |k|,
\eeq%
and, more precisely, when the basic frequency of the acoustic wave
lies within the following interval
\beq\label{rescondac}%
\left | C_{s0} |k| -\frac{\Omega}{2}\right |<\left |
\frac{\mu}{\Omega}\right |,
\eeq%
where $\mu=\delta C_{s0} ^2k^2$. With this notation, the resonant
solution of Eq.~(\ref{waveac}) takes the following form:
\beq\label{ressolac}%
X_{\text{ac}} (t)=X_{\text{ac}}(t=0)%
\exp\left (\frac{\left | \mu \right |}{2\Omega} \tau \right )%
\left [\cos \left (\frac{\Omega}{2}\tau \right )+\sin \left
(\frac{\Omega}{2}\tau \right )\right ]
\eeq%
As a conclusion, when the entropy (temperature) oscillates with a
frequency that is twice the basic frequency of the acoustic wave,
then those waves are resonantly amplified in the
system\footnote{Here, we should emphasize that a similar result
can be obtained when the constraint on the background state given
in the previous section is violated and the density (and hence,
the velocity) variations are also available.}

\subsection{Separation of MHD waves}\label{separation}

Let us now turn to the general analysis of the system
(\ref{wavebasux})-(\ref{wavebasuz}). This system consists of two
wave equations which describe the presence of two different wave
modes. The velocity fields corresponding to these two kinds of
oscillatory motion of the system are clearly coupled.
Nevertheless, if one considers a perturbation around a static
equilibrium state (i.e.\ with $\delta=0$), then it is known that
appropriate eigenvectors can be constructed which correspond to
eigenvalues calculated from the characteristic equation and which
are perpendicular to each other. Using this procedure, one can
then show that the fundamental modes are separated (or decoupled).

\begin{figure*}
  \includegraphics[scale=0.9]{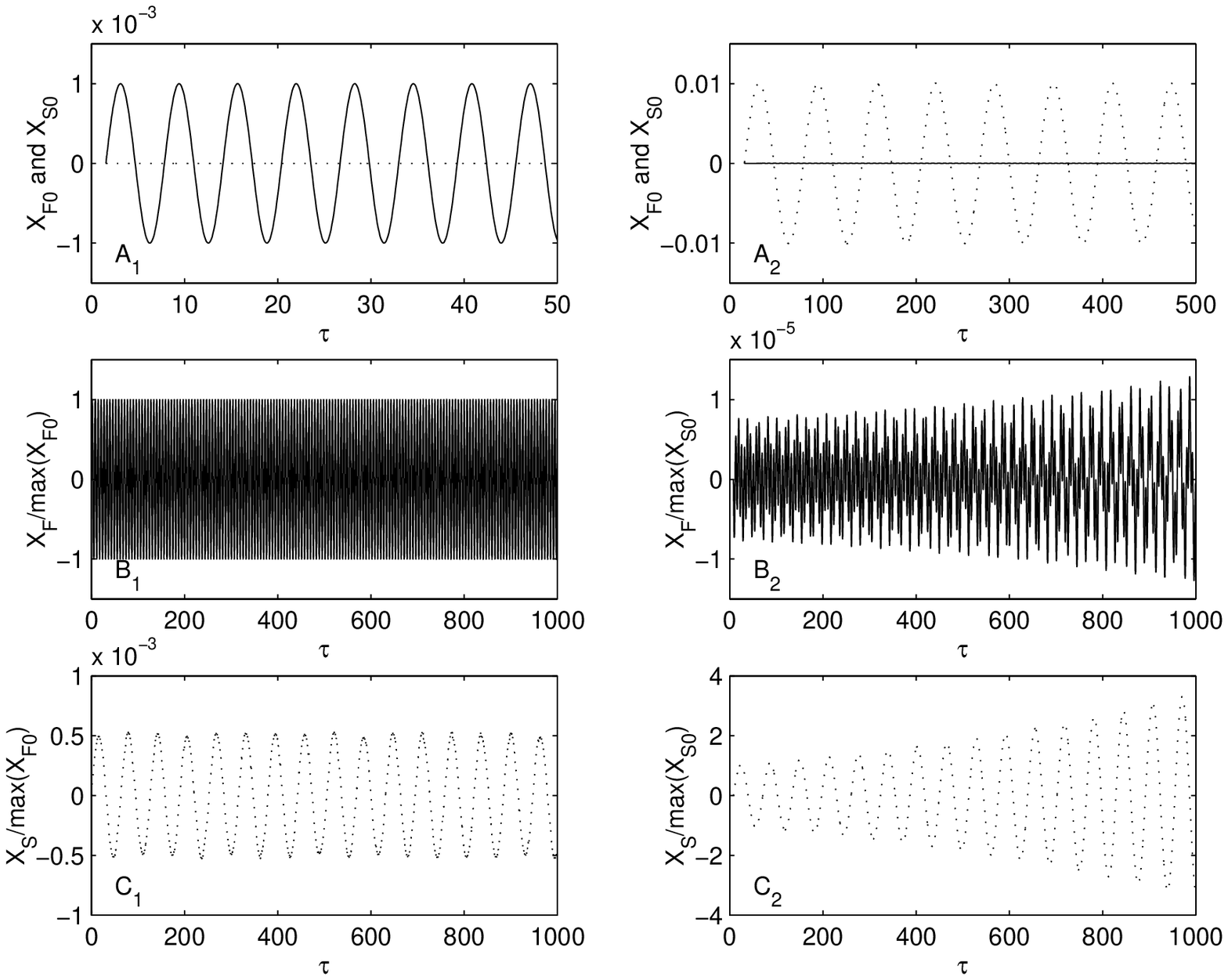}\\
  \caption{\label{figxsmall} As in Fig.~\ref{figxlarge} for $\xi \ll 1$.}
\end{figure*}

However, when the background system is driven (i.e.\ when
$\delta\neq 0$), the coefficients in the equations are time
dependent. Therefore, the analysis of these equations becomes
rather complicated in this case. We are going to construct the
'eigenfunctions' by employing a similar procedure in combination
with numerical techniques, in order to draw a more complete
picture of the dynamical processes we are studying here. For this
purpose it is convenient to impose an appropriate normalization of
the quantities. We introduce the following scaling of the
variables: $\hat{U}_{x}=\hat{u}_{x}/V_A$,
$\hat{U}_{z}=\hat{u}_{z}/V_A$, $\tau=V_A k t$, $C_{11}=c_{11}/V_A
k=\xi ^2K_x ^2+1$, $C_{12}=c_{12}/V_A k=\xi^2K_xK_z$,
$C_{22}=c_{22}/V_A k=\xi ^2K_z ^2$, here, $K_i=k_i/k$ ($i=x,z$),
$\xi^2=C_s^2/V_A ^2=\xi _0 ^2(1+\delta \cos(\hat{\Omega}\tau))$,
and $\hat{\Omega}=\Omega/V_Ak$. Observe that, if we apply an
orthogonal transformation on the velocity vector space
$(\hat{U}_x, \hat{U}_z)$ of the form \cite{courhilb,jeffreys}:
\beq\label{transform}%
\begin{array}{c}
X_{\text{F}}=\cos\theta\hat{U}_x-\sin\theta\hat{U}_z, \\

\\
X_{\text{S}}=\sin\theta\hat{U}_x+\cos\theta\hat{U}_z, \\
\end{array}%
\eeq%
where $\theta$ denotes the Euler angle in the velocity space. This
transformation is time dependent since the coefficients in the
initial matrix:
\beq\label{matrix}%
\left(%
\begin{array}{cc}
  C_{11} & C_{12} \\
  C_{12} & C_{11} \\
\end{array}%
\right)
\eeq%
also depend on time:%
\begin{displaymath}
\tan \theta =\frac{C_{12}}{\Omega _{2}^{2}-C_{11}}=\frac{C_{12}}{%
C_{22}-\Omega _{1}^{2}}=
\end{displaymath} %
\beq%
=\frac{C_{11}-\Omega
_{1}^{2}}{C_{12}}=\frac{\Omega _{2}^{2}-C_{22}}{C_{12}}.%
\eeq%
With this notation, the initial set of equations
(\ref{wavebasux})-(\ref{wavebasuz}) can be transformed into:
\beq\label{waveeqfast}%
\ddot{X}_{\text{F}}+\left( \hat{\Omega} _{\text{F}}^{2}-\dot{\theta}^{2}\right) X_{\text{F}}=\ddot{%
\theta}X_{\text{S}}+2\dot{\theta}\dot{X}_{\text{S}},
\eeq%
\beq\label{waveeqslow}%
\ddot{X}_{\text{S}}+\left( \hat{\Omega} _{\text{S}}^{2}-\dot{\theta}^{2}\right) X_{\text{S}}=-\ddot{%
\theta}X_{\text{F}}-2\dot{\theta}\dot{X}_{\text{F}},
\eeq%
where $\Omega _{\text{F}} ^2$ and $\Omega _{\text{S}} ^2$ are the
eigenvalues of the matrix (\ref{matrix}) representing, in fact, the
characteristic frequencies of the fast (F) and slow (S)
magneto-sonic waves, respectively:
\beq\label{eigenfr}%
\hat{\Omega} _{\text{F},\text{S}}^2=\frac{1}{2}\left
(C_{11}+C_{22} \pm \sqrt{(C_{11}-C_{11})^2+4C_{12}^2} \right ).
\eeq%

It is well known that in a magnetized plasma there exist three
different regimes, which are determined by the relative importance
of the thermal and magnetic pressures.  This relation usually is
parameterized by the ratio of the thermal and magnetic pressures,
the plasma beta defined as $\beta =8\pi p_0/B_0 ^2=\gamma \xi
^2/2$. The proportionality factor here is of the order of unity.
The three regimes are referred to as $\xi^2 \gg 1$, $\xi^2 \ll 1$,
and $\xi^2 \simeq 1$, respectively. Accordingly, the basic
properties of the different MHD waves are also well studied in
these three regimes. Below, we will also consider the problem case
by case. Typically below we use $\xi ^2$ instead of plasma
$\beta$.

\subsection{The case $\xi ^2 \gg 1$}

Firstly, let us consider the limit of large plasma beta. In this
case, the magnetization of the plasma is rather weak and the
thermal effects dominate in the dynamics of the medium. In this
regime, the fast and slow magneto-sonic waves have drastically
different properties. In this situation, one obtains from the
expressions (\ref{eigenfr}) that
\beq\label{lbetaf}%
\hat{\Omega} _{\text{F}} \gtrsim \xi ^2,%
\eeq%
while, on the other hand, the low frequency branch of the spectrum
satisfies:
\beq \label{lbetas}%
\hat{\Omega} _{\text{S}}\lessapprox 1,%
\eeq%
showing that the characteristic frequencies of the slow waves lie
close to the Alfv\'en frequency. In general, the fast and slow
magneto-sonic modes propagating in an arbitrary, oblique direction
with respect to an applied magnetic field are neither purely
compressible (longitudinal) nor incompressible (transversal). The
latter property relates to the tension of magnetic field lines,
like for Alfv\'en waves. But, both of the wave modes consist of a
mixture of these two features. However, when $C_s ^2 \gg V_A ^2$,
the properties of the fast magneto-sonic waves are very similar to
those of sound waves and these waves represent sound waves that
are slightly modified by the presence of the magnetic field and
the transversal component is weak. Therefore, these waves
substantially feel the variation of the sound speed, what directly
results in a variable phase speed of the fast modes.

\begin{figure*}
  \includegraphics[scale=0.9]{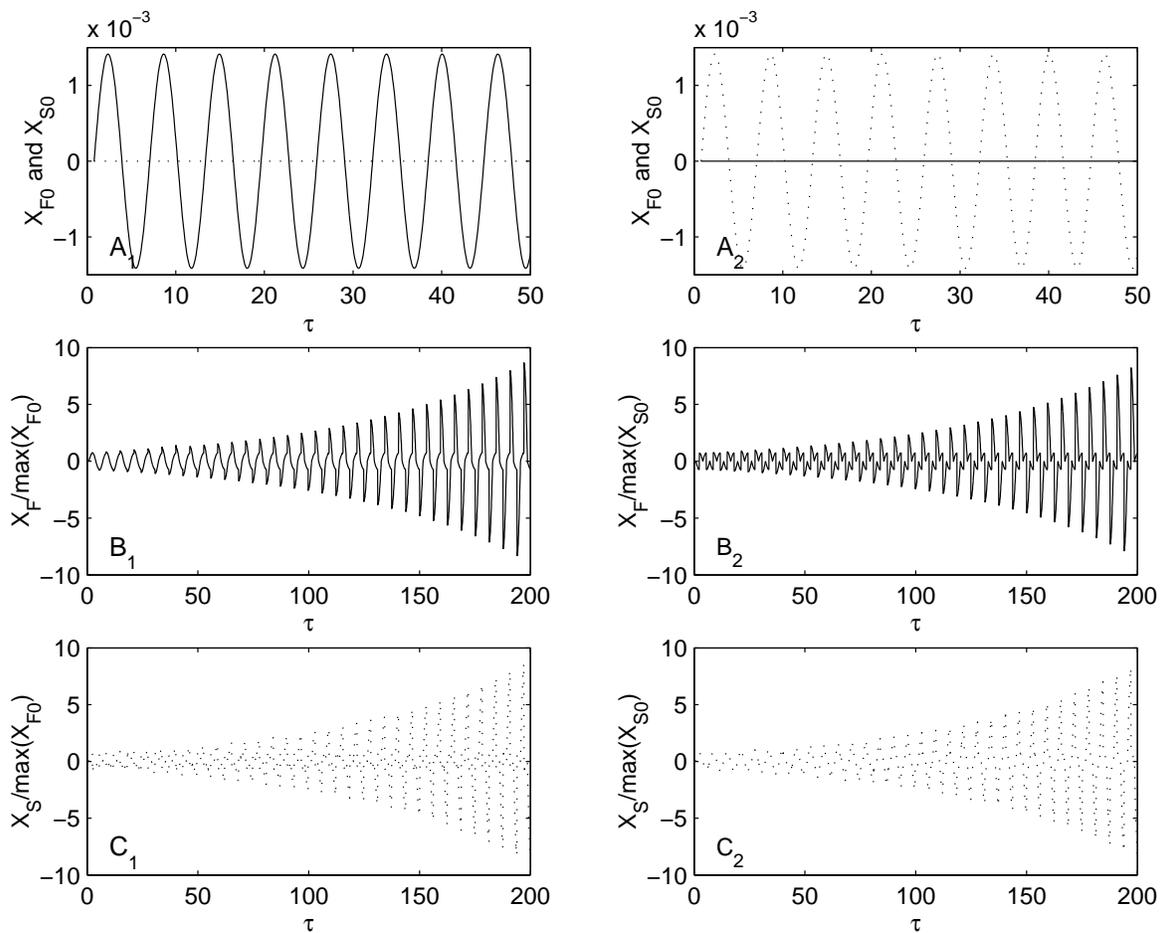}\\
  \caption{\label{figxuni1} As in Fig.~\ref{figxlarge} for $\xi =
  1$. $K_x=0.00005$}
\end{figure*}

We now analyze the set of equations
(\ref{waveeqfast})-(\ref{waveeqslow}) term by term. For this
purpose one should use $\delta$ as a small parameter in the
problem, as we consider here a nonequilibrium background that is
very close to the equilibrium $\delta \ll 1$. We are interested in
terms of the order of magnitude up to linear in $\delta$. All
higher order terms yield vanishingly small contributions to the
considered dynamical process. It can be easily shown that
$\hat{\Omega} _{\text{F}}$ includes both zeroth- and first-order
terms in the expansion in terms of $\delta$, while the terms
$\dot{\theta}$ and $\ddot{\theta}$ do not contain zeroth-order
terms and the expansions of both these terms start with the
first-order term. Consequently, the term $\dot{\theta} ^2$ can be
neglected in both equations as it consists of terms definitely of
second and higher orders in $\delta$. Further, we assume that
within the resonance, the fast magneto-sonic wave would be able to
grow in amplitude exponentially, while the terms controlling the
coupling with the slow waves (on the RHS) are rather small as the
characteristic frequencies of these eigenmodes lie far apart from
each other according to Eqs.~(\ref{lbetaf})-(\ref{lbetas}).
Consequently, we neglect the coupling terms, which at any rate
will be dominated by the terms on the LHS of
Eq.~(\ref{waveeqfast}), and we are left with two approximate \lq
decoupled \rq equations. With these simplifications,
Eq.~(\ref{waveeqfast}) reduces to:
\beq\label{fastmatie}%
\ddot{X}_{\text{F}}+\hat{\Omega} _{\text{F0}}^{2}\left ( 1+
n_{\text{F}} \cos (\Omega \tau +\varphi _{\text{F}})\right )
X_{\text{F}}=0,
\eeq%
where $n_{\text{F}}=n_{\text{F}} (\delta, \Omega)\ll 1$ and
$\varphi _{\text{F}}=\varphi _{\text{F}}(\delta, \Omega)$ are the
amplitude and the initial phase of the fast wave frequency
oscillations in the system, respectively, and $\hat{\Omega}
_{\text{F0}}$ is the equilibrium frequency of the fast
magneto-sonic waves. This Mathieu-type equation has a resonant
solution for the appropriate harmonic of the fast magneto-sonic
wave with the frequency (corresponding to a given value of the
wavelength):
\beq\label{fastcond}%
\hat{\Omega} _{\text{F0}}\approx \frac{\Omega}{2},
\eeq%
varying within the interval
\beq\label{rescondfast}%
\left |  \hat{\Omega} _{\text{F0}}-\frac{\Omega}{2}\right |<\left
| \frac{n_{\text{F}} \hat{\Omega} _{\text{F0}}}{\Omega}\right |,
\eeq%
and reading as:%
\begin{displaymath}
X_{\text{F}} (t)=X_{\text{F}}(\tau=0)%
\exp\left (\frac{\left | n_{\text{F}} \hat{\Omega} _{\text{F0}} ^2
\right |}{2\Omega} \tau \right )\times%
\end{displaymath}
\beq\label{ressolfast}%
\times\left [\cos \left (\frac{\Omega}{2}\tau +\varphi
_{\text{F}}\right )+\sin \left (\frac{\Omega}{2}\tau +\varphi
_{\text{F}}\right )\right ].
\eeq%

The approximative analysis, which is given here, should be
verified by means of experimental or numerical methods. In the
present study, we first concentrate on a rigorous numerical
analysis of the initial equations
(\ref{wavebasux})-(\ref{wavebasuz}), and then we reconstruct from
the results of the numerical simulations, all the relevant
quantities we use in the analysis.

\begin{figure*}
  \includegraphics[scale=0.9]{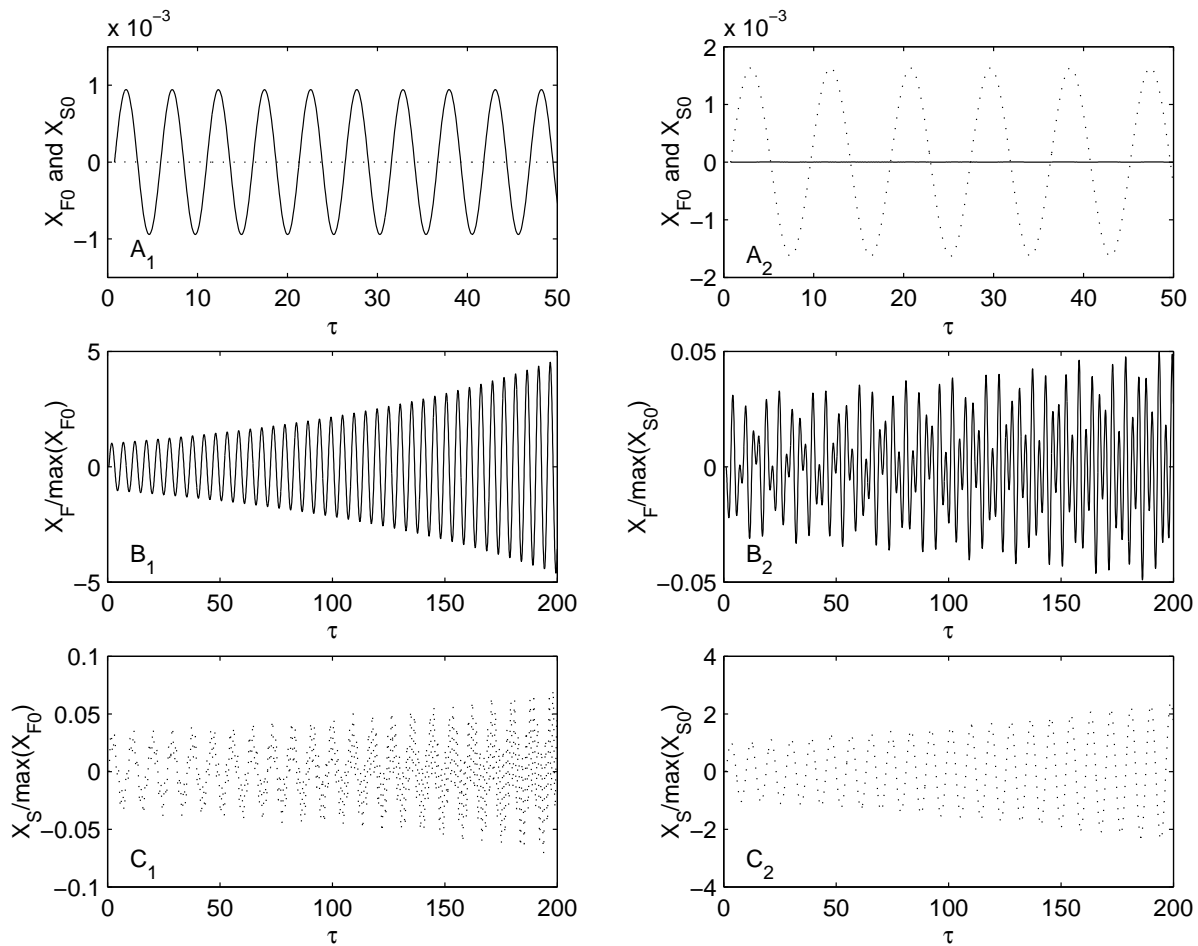}\\
  \caption{\label{figxuni2} As in Fig.~{figxuni1} for $\xi = 1$.
  $K_x=0.5$}
\end{figure*}

Here, we consider two different sets for the initial conditions.
First we chose the initial conditions $\hat{U}_{x0}$,
$\hat{\dot{U}}_{x0}$, $\hat{U}_{z0}$ and $\hat{\dot{U}}_{z0}$ so
that, initially, only the amplitude of the fast magneto-sonic wave
(solid line in panel $A_1$) is finite and there is no slow
magneto-sonic wave in the system (see the dotted line in panel
$A_1$):
\beq\label{initfast}%
X_{\text{S}}(\tau=0)\sim \varepsilon, \hskip 0.3 cm
\dot{X}_{\text{S}}(\tau=0)\sim \varepsilon,
\eeq%
where $\varepsilon\approx 10^{-17}$ is the order of magnitude of
the numerical error related to the machine precision. Panel $A_1$
represents the case of the stationary state in the system
($\delta=0$). On the other hand, in panels $B_1$ and $C_1$ of the
same figure, the curves corresponding to the parameter values
$\delta=0.05$ and $\hat{\Omega} _{\text{F0}}=\Omega/2\approx 10$
are plotted. It is obvious that the amplitude of the fast
oscillation undergoes a rapid exponential growth in this case
(panel $B_1$, Fig~\ref{figxlarge}), while the slow magneto-sonic
wave is out of resonance and, consequently, its amplitude remains
small (panel $C_1$). The excitation of this small amplitude slow
magneto-sonic disturbance is due to the weak coupling of the fast
and slow waves by the external driver. This coupling implies that,
even when there is initially no slow magneto-sonic wave,
nevertheless, a small amplitude slow magneto-sonic wave
perturbation gets excited. These small amplitude vibrations can be
referred to as {\lq}slow{\rq} magneto-sonic noise. The presence of
this noise relates to two simultaneous factors, viz.\ the
propagation of the fast magneto-sonic wave and the external
driving of the background system. The exclusion of either of these
two factors would lead to the disappearance of the slow
magneto-sonic perturbation, as is shown in panel $A_1$.

On the other hand, we have carried out the simulations for the
case when frequency of slow waves satisfies the resonant condition
$\hat{\Omega} _{\text{S0}}=\Omega/2\approx 1$. The sound speed
yields a negligible contribution into the frequency of the slow
magneto-sonic waves, as from expression (\ref{lbetas}) it is
evident that the latter is close to the Alfv\'en frequency (which
in our case is constant), and whatever variations of the sound
speed (related to the pressure variation) occur in the system, the
slow magneto-sonic wave frequency remains almost unchanged. The
corresponding values of the increment $\delta _S$ is so small that
there is not a significant growth of the slow mode amplitude
during any reasonable time span. This means that this variation is
not felt by slow magneto-sonic wave modes. In the slow
magneto-sonic waves propagating in a high-beta plasma, the
incompressible (transversal) component dominates and the
properties of these waves are very similar to the purely
incompressible Alfv\'en waves. Therefore, one can expect that the
slow magneto-sonic waves (and also the Alfv\'en waves) do not feel
the variations of the sound speed when the Alfv\'en speed remains
constant. This is why, in the case when we initially have only the
slow magneto-sonic wave ($A_2$ of Fig.~\ref{figxlarge}), these
waves stay out of the parametric interaction (panel $C_2$), even
when an appropriate resonant conditions is formally satisfied. In
panel $B_2$ of Fig.~\ref{figxlarge}, we plot $X_{\text{F}}$
showing that the fast magneto-sonic wave naturally is out of
resonance due to the fact that its frequency is out of the
resonant area. Again, the presence of the slow magneto-sonic wave
of finite amplitude in the externally driven system leads to the
excitation of small amplitude disturbances (panel $B_2$), which we
can now refer to as {\lq}fast{\rq} magneto-sonic noise.

To involve the almost incompressible slow magneto-sonic waves into
the parametric interaction, one should not consider the case of
constant density (for the parametric amplification of Alfv\'en
waves with periodical variation of medium density see
\cite{zaq2001}). Instead of Eq.~(\ref{entpert2}), the full
equation Eq.~(\ref{entpert}) should then be employed. In that
case, the action of the external periodic forces can also be taken
into account. However, along with the effect of parametric
resonance, we expect that strong couplings between the
compressional fast and slow and incompressible Alfv\'en waves can
arise. Therefore, the problem would then not be reducible to two
spatial dimensions any more, since the Alfv\'en speed also would
be variable and the Alfv\'en waves would participate in the
process as well. This issue can become the subject of future
developments and generalizations of the present model.

\subsection{The case $\xi ^2\ll 1$}

In the present subsection, we address the case when the plasma is
magnetically dominated, i.e.,\ the case $C_s ^2 \ll V_A ^2$. In
this regime, the situation is just opposite compared to the
previous one. In particular, we know that now the frequency of the
fast magneto-sonic wave satisfies
\beq\label{sbetaf}%
\hat{\Omega} _{\text{F}} \gtrsim 1,%
\eeq%
and the properties of these waves are close to those of the
incompressible Alfv\'en waves. Consequently, in a low-beta plasma
the fast magneto-sonic waves almost do not feel the variation of
the entropy. However,  this is again true only when an external
driver causes a variation of only the pressure while the density
remains constant. In general, when the Eq.~(\ref{entpert}) holds
so that the density is variable, the remark given in the previous
subsection about the slow magneto-sonic waves and Alfv\'en waves
is now valid for fast magneto-sonic waves.

Within the present framework, the fast magneto-sonic waves stay
out of resonance. This can also be discovered by means of
numerical simulations. In Fig.~\ref{figxsmall}, we show the
results of our simulations for the new set of parameter values in
the same order as in Fig~\ref{figxlarge}. Panel $B_1$ confirms the
validity of the conclusions we have just made.

However, one can again follow an approximative analysis for the
slow magneto-sonic waves, in a similar manner as in the previous
case. Indeed, contrary to the case of high plasma beta, now the
slow magneto-sonic modes are predominantly compressional
(longitudinal) and their properties are close to those of the
usual sound waves:
\beq\label{sbetas}%
\hat{\Omega} _{\text{S}} \lesssim \xi ^2.%
\eeq%
Therefore, following a similar logic as before, one can derive an
approximate Mathieu-type equation for the slow waves which reads:
\beq\label{slowmatie}%
\ddot{X}_{\text{S}}+\hat{\Omega} _{\text{S0}}^{2}\left ( 1+
n_{\text{S}} \cos (\Omega \tau +\varphi _{\text{S}})\right )
X_{\text{S}}=0,
\eeq%
where $n_{\text{S}}=n_{\text{S}} (\delta, \Omega)\ll 1$ and $\varphi
_{\text{S}}=\varphi _{\text{F}}(\delta, \Omega)$, respectively, are
the amplitude and the initial phase of the slow magneto-sonic wave
frequency oscillations in the system, and $\hat{\Omega}_{\text{S0}}$
is the equilibrium frequency of the slow magneto-sonic waves. Again,
this Mathieu-type equation has a resonant solution for the
appropriate harmonic of the slow magneto-sonic wave with the
frequency (corresponding to a given value of the wavelength):
\beq\label{slowcond}%
\hat{\Omega} _{\text{S0}}\approx \frac{\Omega}{2},
\eeq%
varying within the interval
\beq\label{rescondslow}%
\left |  \hat{\Omega} _{\text{F0}}-\frac{\Omega}{2}\right |<\left
| \frac{n_{\text{F}} \hat{\Omega} _{\text{F0}}}{\Omega}\right |,
\eeq%
and it has the form:
\begin{displaymath}
X_{\text{S}} (t)=X_{\text{S}}(\tau=0)%
\exp\left (\frac{\left | n_{\text{S}} \hat{\Omega} _{\text{S0}}^2
\right |}{2\Omega} \tau \right )\times%
\end{displaymath}
\beq\label{ressolslow}%
\times \left [\cos \left (\frac{\Omega}{2}\tau +\varphi
_{\text{S}}\right )+\sin \left (\frac{\Omega}{2}\tau +\varphi
_{\text{S}}\right )\right ]
\eeq%
Furthermore, we verify these findings by a numerical simulation
when $\hat{\Omega} _{\text{S0}}=\Omega/2\approx 0.1$. As we show
in panel $C_2$ of Fig.~\ref{figxsmall}, the slow magneto-sonic
waves get amplified, while the fast magneto-sonic waves stay out
of resonance (panel $B_2$), and the coupling between these waves
is again very weak because $\hat{\Omega} _{\text{S0}}\ll
\hat{\Omega} _{\text{F0}}$. Observe that in both cases of the
initial conditions (panels $A_1$ and $A_2$), we can see the
excitation of the small amplitude {\lq}slow{\rq} magneto-sonic
(panel $C_1$) and {\lq}fast{\rq} magneto-sonic (panel $B_2$)
noises due to the weak coupling of the waves.

\subsection{The case $\xi \approx 1$}
This case is of special importance in the problem considered here.
When $C_s ^2\approx V_A^2$, the thermal and magnetic effects are of
comparable strength. Under these conditions, the properties of both
waves are very similar as:
\beq\label{unibetas}%
\hat{\Omega} _{\text{F}} \approx \hat{\Omega} _{\text{S}}.
\eeq%
That is the reason why the waves can not only interact with the
forced entropy oscillations but, in addition, they also may
effectively couple and exchange energy among each other. This can be
understood easily if we recall the presence of the terms in the RHS
of Eqs.~(\ref{waveeqfast})-(\ref{waveeqslow}). These equations show
that an external driver can effectively couple the waves. This
situation is similar to the case of a laminar shear flow in the
system \cite{gogo2004}. As is well-known, the latter also can be
maintained only by an external driving of the system. This seems to
be a general property of driven systems: the ability to couple
eigenmodes of the system under certain conditions. This claim looks
apparent in the framework of the current paper. However, a complete
and general proof of it can be given only through the general
consideration of the fully three-dimensional case when the
variability of the background density and the action of the external
forces also are taken into account. Here, we should emphasize that
the coupling of different kinds of wave modes can be achieved not
only with the external driving, which harmonically oscillates in
time, but in general, with any kind of driver giving rise to a time
dependent entropy (temperature) in the system.

Now, a similar approximative analysis as we have discussed in the
previous two cases, would be very {\lq}rough{\rq}. In the case
when the waves propagate almost along the magnetic field ($K_x \ll
1$), it even becomes completely inapplicable. Consequently, it is
convenient to treat the problem in this case only numerically.
Before we do this, it is important to emphasize again that the
coupling of the different wave modes is possible only because of
the presence of the external driving in the system. The
simulations show that the dynamics of both wave modes (when the
resonant conditions are satisfied) is characterized by the
interplay of two different factors. On the one hand, both wave
modes are involved in the parametric resonance and, on the other
hand, they are effectively coupled to each other.

Let us first consider the case when both fast and slow
magneto-sonic modes propagate almost parallel to the magnetic
field ($K_x \ll 1$). In this case, in \ref{figxuni1} we
accordingly show the results of the simulations as in the previous
subsections. It is apparent that both {\lq}eigenfrequencies{\rq}
lie very close to each other. As a consequence, in
Fig.~\ref{figxuni1} we show that, if initially we have only a fast
magneto-sonic wave (panel $A_1$) and the resonant condition is
satisfied ($\hat{\Omega}=2$), the fast magneto-sonic mode gets
amplified via the parametric action (panel $B_1$). Besides, the
coupling of the two wave modes is very strong in this case and,
therefore, the slow magneto-sonic wave is excited and also becomes
involved in the amplification process, as this wave satisfies the
resonance condition as well (panel $C_1$). Both kinds of wave
modes effectively exchange energy and the characteristic values of
their amplitudes are comparable to each other. The situation is
absolutely similar, when initially we set up only a slow
magneto-sonic wave (panel $A_2$). We then observe that a fast
magneto-sonic wave is excited (panel $B_2$) and both waves are in
parametric resonance with the driver as they exchange energy
(panels $B_2$ and $C_2$). It is important to note that, in this
case, the ultimate {\lq}fate{\rq} of the process is more sensitive
to the initial conditions as, contrary to the case of weak
coupling, not only the frequencies are important but also their
phases matter.

We also investigated the case when the waves propagate in the
oblique direction $K_x=0.5$. In this case, the frequencies of the
modes become somewhat separated even both of them still are of
same order of magnitude. Therefore, the coupling effect weakens
substantially (as shown in Fig.~\ref{figxuni2}), and the
parametric resonance effect only effectively amplifies, with the
respective resonant conditions, in the one case only the fast
magneto-sonic wave (panel $B_1$) and, in the other case, only the
slow magneto-sonic wave (panel $C_2$). At the same time, it should
be noticed that again we discover the generation of the
{\lq}slow{\rq} (panel $C_1$) and {\lq}fast{\rq} (panel $B_2$)
magneto-sonic noises, which are of much larger amplitude now (as
expected) than in the limits of high or low plasma beta.

\section{Conclusions}\label{concl}

In the present paper we investigated wave motions and wave
amplifications in a non-equilibrium setting. In particular, we
have developed a model of a magnetohydrodynamic system that is
being driven by the combined action of external sources and sinks
of heat.  More specifically, the external action is a harmonically
oscillating function of time. In contrast with many other studies,
the background is that of a non-equilibrium state against which we
have considered the dynamical properties of the compressional fast
and slow magneto-sonic waves. Of special interest has been the
case when the external driving of the system results only in a
variation of the pressure (or temperature) and it does not lead to
a distortion of the background velocity and density fields. A
numerical study has led us to the following conclusions:
\begin{enumerate}
\item In the hydrodynamic limit ($B_0=0$), the spectrum of the
possible wave modes reduces to that of the sound waves. In this
case, we derived an exact equation, viz.\ Eq.~(\ref{waveac}) with an
obvious resonant solution given by Eq.~(\ref{ressolac}).

\item In the case of a high beta plasma ($\xi ^2 \gg 1$), the
temperature oscillation leads to a major contribution in the
variability of the fast magneto-sonic wave frequency $\hat{\Omega}
_{\text{F}}$, while the amplitude of the slow magneto-sonic
frequency $\hat{\Omega} _{\text{S}}$ oscillation  is very small.
As a consequence, in this regime only the fast magneto-sonic waves
can participate in the parametric resonance with the entropy
oscillations (panel $B_1$ in Fig.~\ref{figxlarge}) as is predicted
by the approximate solution (\ref{ressolfast}). The slow
magneto-sonic waves stay out of resonance because the background
density (i.e. Alfv\'en speed) is assumed to be constant (cf.\
panel $C_2$ in Fig.~\ref{figxlarge}), even when the appropriate
condition for the parametric resonance is satisfied. Besides,
there is a coupling between the fast and slow magneto-sonic waves
due to the open nature of the system, as is manifested from the
terms on the RHS of Eqs.~(\ref{waveeqfast})-(\ref{waveeqslow}).
This coupling is very weak when the plasma $\beta$ is very large
and the frequencies of the waves are very separated. Nonetheless,
this coupling leads to the excitation of slow (cf.\ panel $C_1$ in
Fig.~\ref{figxlarge}) and fast (cf.\ panel $B_2$ in
Fig.~\ref{figxlarge}) ``noises'' of very small amplitude.

\item The situation is just opposite in the case of low
plasma-$\beta$ ($\xi ^2 \ll 1$). In this case only the slow
magneto-sonic wave frequency oscillates with the significant
amplitude, while the variability of the fast magneto-sonic wave
frequency is vanishingly small. Hence, only slow waves can be
amplified (cf.\ panel $C_2$ in Fig.~\ref{figxsmall}) by the
parametric action of the forced entropy oscillations and the fast
magneto-sonic modes do not feel (in the sense of parametric
action) these oscillations effectively (cf.\ panel $B_1$ in
Fig.~\ref{figxsmall}) with the constant Alfv\'en speed considered
in current work. In addition, we again observe the excitation of
slow magneto-sonic (see panel $C_1$ in Fig.~\ref{figxsmall}) and
fast magneto-sonic (see panel $B_2$ in Fig.~\ref{figxsmall})
``noises'' due to the weak mode coupling.

\item When ($\xi ^2 \sim 1$), we can distinguish two different
situations:

(i)~In the case when both fast and slow magneto-sonic waves
propagate almost parallel to the magnetic field $K_x \ll 1$, the
characteristic frequencies of the different modes are very close
to each other. Therefore, there is a strong wave coupling. As a
result, there is no possibility for only one kind of wave mode to
exist, \ie the presence of one of the wave modes immediately
results in the excitation of the other. In this case, both kinds
of waves can be involved together in the parametric resonance as
shown in Fig.~\ref{figxuni1}.

(ii) However, when the waves propagate in the oblique direction
with respect to the applied magnetic field, the frequencies become
somewhat more separated and then the coupling effect weakens
significantly (Fig.~\ref{figxuni2}).
\end{enumerate}

The results obtained in this work can be applied  under
terrestrial laboratory conditions such as in acoustic
investigations but the model also appears useful for the
description of some astrophysical processes that are periodic in
time, as in objects with a variable intensity of entropy
production and losses (as for instance in variable stars).
However, at the present stage we aim only to demonstrate the basic
physical properties of the parametric amplification of waves and
their mutual coupling in  nonequilibrium plasmas. Despite of the
apparent simplifications our new results are believed to be of
general importance and they may have far reaching consequences for
many applications. Here follow some possibilities:

\begin{itemize}
\item Previous investigations of processes based on parametric
resonance in neutral and in conducting fluids have been carried
out. These studies have revealed the possibility of an effective
parametric amplification of  sound waves either due to external
mechanical periodic forcing or to mutual nonlinear interactions
between the different harmonics of the waves themselves. Along
with the references on this matter given in the previous sections,
other examples include \cite{silva2006,stern1966,tokman2005}. The
model outlined in the present work is a unified approach to
parametric amplification of sound and of compressional MHD waves
in considered two dimensional systems. The generalization to 3D
would also enable an experimental treatment of the parametric
resonance triggered by an externally driven, oscillating entropy
(or temperature). Thus, it would yield a valuable contribution to
the rich spectrum of theoretical and experimental studies (for
instance, see \cite{rosenbluth1972,sugawa1989} and many others) in
the field of parametric instabilities in fusion plasmas.

\item The spectrum of astrophysical processes, where the
considered mechanism of wave amplification and of wave coupling
may be useful is even more rich. We have already given (see above)
several references concerning the swing (parametric) interactions
between different MHD waves in equilibrium magnetic structures,
which directly have been applied to the magnetized solar
atmosphere. On the other hand, there are many studies, which
reveal the importance of parametric resonances at galactic and
cosmological scales (for instance, see
\cite{easther2000,henriques2002,servin2000,zibin2001}). It should
be emphasized that the results obtained in this work may explain
the properties of oscillatory phenomena in astrophysical objects
being away from thermodynamic equilibrium and sustaining forced
entropy oscillations. The entire variety of such thermally
variable objects are usually  described within the framework of
thermal relaxation oscillation theory \cite{wesse1939}. We give
here references to types of the variable stars which manifest the
mentioned oscillatory behavior: Thermal oscillation during the
carbon burning  \cite{iben1978} and Helium burning \cite{iben1986}
in the stellar core; Also thermal relaxation in the
horizontal-branch stars \cite{salati1990} and those with
convective cores \cite{wood2000}. In addition, our approach
appears interesting for models of globally oscillating coronas
\cite{korevaar1989}, although such a consideration would require a
generalization of the current model for the case when the
variability of the density is also  taken into account. In such
systems the parametric amplification of the physical quantities
could even become observable if it would be possible to observe
two, mutually related oscillatory processes with the frequencies
related as $\omega _1 \approx 2\omega _2$.
\end{itemize}

The model presented here, of course, requires some further
development, e.g.\ to enable the study of similar effects when the
background density and velocity are also variable in time. In
addition, the terms corresponding to a perturbation of the source
term in the entropy balance equation, due to the propagation of
waves, could also be taken into account (in the current model we
neglected this term). In that case, additional kinds of modes can
arise in the system such as thermal waves and aperiodic vortices
which would enrich the spectrum of possible interactions.

\begin{acknowledgements}
These results were obtained in the framework of the projects
GOA/2004/01 (K.U.Leuven), G.0304.07 (FWO-Vlaanderen) and
C~90203 (ESA Prodex 8). 
The work have been partially supported by K.U.Leuven scholarship
-PDM/06/116 and Grant of Georgian National Science Foundation -
GNSF/ST06/4-098. We are thankful to the anonymous referee for
constructive comments on our paper.
\end{acknowledgements}
\appendix*
\section{Relation of the heat conduction with the entropy production and its current}
The current derivation is based on standard linear response theory
for weakly non-equilibrium systems {e.g.\ see \cite{grootmaz}}.
Consider the source term $\lambda \Delta T$ in
Eq.~(\ref{fouriertemp}). It is well known that the coefficient of
thermal conduction $\lambda $ arises here due to the assumption
that it depends only on the overall equilibrium temperature
$\lambda =L_{qq}/T_{00}^{2}$, where $L_{qq}$ is a constant
parameter. Setting $L_{qq}=0$ in the considered case reduces the
system to the equilibrium. On the other hand, the assumption that
the coefficient $\lambda $ is a constant is known as the Fourier
approximation, leading to the Fourier law of heat transfer
(\ref{fouriertemp}). From a general point of view the equation
which governs the heat transport reads as:
$$
\rho _0 T_0\frac{D S_0}{D t}=\mathbf{\nabla}\cdot \left( \frac{L_{qq}}{T_0^{2}}\mathbf{%
\nabla}T_0\right) =-\mathbf{\nabla}\cdot \left( L_{qq}\mathbf{\nabla}\frac{1}{T_0}%
\right) =-\mathbf{\nabla}\cdot \mathbf{j}_{q},
$$
which says that the changes in the local entropy are caused by the
heat transfer through the system (if there is no source of heat
within it). The divergence of the heat current on the r.h.s.\ can
be reorganized as follows:
\begin{equation}\label{reorglamb}
-\mathbf{\nabla}\cdot \mathbf{j}_{q}=T_0\left( -\mathbf{\nabla}\cdot \mathbf{j}%
_{s}+\sigma \right),
\end{equation}
which recovers the  expression (\ref{source}) with the microscopic
entropy current $\mathbf{j}_{s}=\mathbf{j}_{q}/T_0$ and local
entropy production
\beq\label{entpro}%
\sigma =\mathbf{j}_{q}\cdot \mathbf{\nabla}\left(
\frac{1}{T_0}\right).
\eeq%
These last equalities reveal the relation between the coefficient
of thermal conduction $\lambda$, and the entropy current and its
production. As a matter of fact, the presence of the coefficient
$\lambda $ in Eq.~(\ref{fouriertemp}) manifests the presence of
both entropy production and its transfer, thus determining the
system to be away form thermodynamic equilibrium.

Further, in the Fourier approximation one can write:
\beq\label{entap}%
\sigma _{0}\approx \frac{\lambda ^{2}}{L_{qq}}\left( \mathbf{\nabla}%
T_{0}\right) ^{2}
\eeq%
\beq\label{entcurap}%
-\mathbf{\nabla}\cdot \mathbf{j}_{s}\approx \frac{\lambda \Delta T_{0}}{T_{0}}-%
\frac{\lambda ^{2}}{L_{qq}}\left( \mathbf{\nabla}T_{0}\right) ^{2}
\eeq%
These expression can be written explicitly for the example
(\ref{tempmode}):
\beq\label{entapsol}%
\sigma _{0}\approx \frac{\lambda ^{2} T_{01}^{2}}{L_{qq}}%
f^{2}\exp \left( 2fx\right) \exp (2i\omega t),
\eeq%

\begin{displaymath}%
-\mathbf{\nabla}\cdot \mathbf{j}_{s}\approx -\frac{\lambda ^{2}
T_{01}^{2}}{L_{qq}}f^{2}\exp \left( 2fx\right) \exp (2i\omega t) +
\end{displaymath}
\beq\label{entcurapsol}%
+\frac{\lambda f^{2}T_{01}\exp \left( fx\right) \exp (i\omega t)}{%
T_{00}+T_{01}\exp \left( fx\right) \exp (i\omega t)},
\eeq%
where,
\beq%
 f=\left( -\sqrt{\frac{\omega \rho C_{v}}{2\lambda
}}(1-i)\right).
\eeq%
 As it has been mentioned in the text we omit the coordinate
dependence of these quantities implying that $\exp \left(
fx\right) \approx \exp \left( fx_{0}\right)$, where $x_0$ is a
constant. This last assumption immediately leads to an equation of
type (\ref{entpert2}) with a constant amplitude $\alpha_1$, which
is under consideration here.


\end{document}